\documentclass[twocolumn]{aastex63}

\newcommand\degree{^{\circ}}
\usepackage{wasysym}
\usepackage{colortbl}

\definecolor{mgm}{rgb}{0.0, 0.0, 1.0}

\definecolor{trq}{rgb}{0.4, 0.8, 0.3}

\begin{document}

\shorttitle{Resonance in Three Systems}
\shortauthors{Quinn and MacDonald}

\email{macdonam@tcnj.edu}

\title{Confirming Resonance in Three Transiting Systems}
\author[0000-0002-8974-8095]{Tyler Quinn}
\affiliation{Department of Astronomy \& Astrophysics, Center for Exoplanets and Habitable Worlds, The Pennsylvania State University, University Park, PA 16802, USA}
\author[0000-0003-2372-1364]{Mariah G. MacDonald}
\affiliation{Department of Astronomy \& Astrophysics, Center for Exoplanets and Habitable Worlds, The Pennsylvania State University, University Park, PA 16802, USA}
\affiliation{Department of Physics, The College of New Jersey, 2000 Pennington Road, Ewing, NJ 08628, USA}

\begin{abstract}

Although resonant planets have orbital periods near commensurability, resonance is also dictated by other factors, such as the planets' eccentricities and masses, and therefore must be confirmed through a study of the system's dynamics. Here, we perform such a study for five multi-planet systems: Kepler-226, Kepler-254, Kepler-363, Kepler-1542, and K2-32. For each system, we run a suite of \textit{N}-body simulations that span the full parameter-space that is consistent with the constrained orbital and planetary properties. We study the stability of each system and look for resonances based on the libration of the critical resonant angles. We find strong evidence for a two-body resonance in each system;   we confirm a 3:2 resonance between Kepler-226c and Kepler-226d, confirm a 3:2 resonance between Kepler-254c and Kepler-254d, and confirm a three-body 1:2:3 resonant chain between the three planets of Kepler-363. We explore the dynamical history of two of these systems and find that these resonances most likely formed without migration. Migration leads to the libration of the three-body resonant angle, but these angles circulate in both Kepler-254 and Kepler-363. Applying our methods to additional near-resonant systems could help us identify which systems are truly resonant or non-resonant and which systems require additional follow-up analysis.

\end{abstract}



\section{Introduction} \label{sec:intro}

While in operation, the Kepler space telescope discovered over 4,500 planet candidates during both the Kepler and K2 missions. Today, many of these candidates have been confirmed, and Kepler-era exoplanets have contributed to the growth of the confirmed exoplanet catalog to over 5,000 and the catalog of candidate planets to over 8,500. This large sample size has led to many investigations into planetary composition, formation, dynamics, and evolution through astrobiological studies.

One intrigue raised by these studies is mean-motion resonance (MMR). MMR occurs when two or more orbiting bodies periodically exert gravitational perturbations on each other, leading to a repeated exchange of energy and angular momentum. We can predict MMR by observing the orbital frequency of neighboring planets. If in resonance, the ratio of neighboring planets’ periods will reduce to a ratio of small integers, such as 2:1 or 12:5. However, determining resonance requires a deeper study into the system’s dynamics since a period ratio of small integers does not necessarily mean the system is in resonance. Such in-depth studies have been conducted and confirmed resonance in a handful of Kepler systems such as Kepler-80 \citep{MacDonald_2016}, Kepler-223 \citep{mills2016resonant}, and K2-138 \citep{MacDonald_2022}.

Mean-motion resonance can form in systems with two or more orbiting bodies. The simplest form of MMR is the two-body resonance. 
Mathematically, this is defined as the oscillation or libration of the two-body critical angle: \begin{equation}\label{twobodyres}
\Theta_{b,c} = j_1\lambda_b + j_2\lambda_c + j_3\omega_b + j_4\omega_c + j_5\Omega_b + j_6\Omega_c
\end{equation}
\noindent where $\lambda_p$ is the mean longitude of planet $p$, $\omega_p$ is the argument of periapsis, $\Omega_p$ is the longitude of the ascending node, $j_i$ are coefficients which sum to zero, and planet $b$ orbits closer to the host star than planet $c$.

In systems with three or more orbiting bodies, numerous bodies may be in resonance, either in a chain of two-body resonances or in a three or more body resonance. A zeroth-order three-body MMR is defined by the difference of the two-body resonant angles: \begin{equation}\label{threebodyres}
    \phi_{b,c,d} = \Theta_{c,d} - \Theta_{b,c} = m\lambda_d - (m + n)\lambda_c + n\lambda_b
\end{equation}
\noindent where $\lambda_p$ is the mean longitude of planet $p$, and $m$ and $n$ are integers. This angle is independent of all longitudes of periapsis ($\bar{\omega}=\Omega+\omega$), making it ideal for resonant study in systems with poorly constrained orbital angles and eccentricity.

Traditionally, such resonances are confirmed if all solutions to the system's RV or TTV forward modeling lead to librating angles. Unfortunately, few systems produce large enough perturbations that could be detected  with a typical survey cadence (30 minute cadence from photometry and $\sim$few day cadence from radial velocities). Due to a lack of high-precision measurements of these systems, we must model all solutions to a system---across all potential parameters that are consistent with the data---to confirm resonance. 
In the case that all solutions result in the planets locked in MMRs, we are able to confirm resonance in the system.

\citet{MacDonald_2022} were the first to confirm a resonance without forward modeling either transit times or the radial velocity signal of the planets. They found that three of the planets of K2-138 are locked in a resonant chain in 99\% of $N$-body simulations that spanned the entirety of parameter space that was previously constrained by both photometry and radial velocity measurements, providing a method of MMR confirmation in the absence of high-cadence, high-precision data.

Such a method, if applied on a larger scale to more systems, would enable us to confirm more resonances. Since resonances allow for the constraint of planetary properties, the system's formation history, and the planets' long term stability, a significant number of confirmed resonant systems would allow us to start leveraging these dynamics to better understand planet formation and evolution. 

Here, we perform such an analysis on five systems: Kepler-226, Kepler-254, Kepler-363, Kepler-1542, and K2-32. Each of these systems was suggested to be a ``broken,'' full-system 3:2 resonant chain, where the discovery of an additional planet would complete the chain \citep{christiansen2018k2}. However, the period ratios of adjacent known planets suggest the presence of resonant chains.  Very few known systems with similar architecture exist 
\citep{2018livingston}, and confirmation of such a chain can provide valuable insight into the dynamics, history, and composition of systems of this architecture.

In Section~\ref{sec:background}, we briefly describe the five systems we study  and discuss the initial conditions and parameters of our \textit{N}-body simulations. We then present our results and analyze the resonant configurations of each system in Section~\ref{sec:Results}. For two of the systems in which we confirm resonance, we use the resonances to constrain the planetary masses and orbital periods and discuss forming the chain in Section~\ref{sec:discussion} before summarizing and concluding our work in Section~\ref{sec:conclusion}.


\section{Methods}\label{sec:background}

Kepler-226 is a G-type star hosting a super-Earth and two Earth-sized planets with orbital periods between 4 and 8 days. These three planets could be locked in a 2:3:4 resonant chain. Since their initial confirmation \citep{rowe}, the anti-correlated TTVs of planets b and c constrained their masses to $M_b=24.0^{+11.8}_{-10.1}~M_{\oplus}$ and $M_c=45.2^{+22.5}_{-19.1}~M_{\oplus}$, although the radii of these two planets \citep[$R_b=1.64~R_{\oplus}$ and $R_c=2.47~R_{\oplus}$,][]{berger2018kepler} suggest these values to be overestimates. Although the TTVs and period ratios of the system suggest this chain of resonances, the specific dynamics of the system have yet to be explore.

Kepler-254 is a relatively dim ($V=16.012$) G-type star, hosting three confirmed exoplanets with orbital periods ranging from 5.8 days to 18.7 days. The period ratios of adjacent planets suggest the system could be locked in a 1:2:3 resonant chain. \citet{2021hutter} suggest that Kepler-254d and Kepler-254c could be locked in a 3:2 resonance. However, the orbital dynamics of Kepler-254 have yet to be included in an in-depth study to confirm MMRs. 

Kepler-363 is a relatively bright ($V=13.472$) G-type star, hosting three confirmed exoplanets. These planets orbit their stair fairly rapidly, with orbital periods ranging from 3.6 days to 11.9 days. The period ratios of adjacent planets suggest the system could be locked into a 1:2:3 resonant chain. The orbital dynamics of Kepler-363 have yet to be included in any in-depth study to confirm resonance in the system.

Kepler-1542 is a G-type star that hosts four transiting planets and one planetary candidate, all smaller than Earth and orbiting within 8 days. The orbital periods of the planets suggest a chain of resonances of 4:3, 5:4, 7:6, and 6:5 if we include the candidate. Validated by \citet{morton2016kepler}, the four planets have never been included in an in-depth study of the system.

K2-32 is a G-type star in a binary system, hosting four transiting planets. The innermost planet K2-32e was most recently discovered and validated by \citet{Heller2019}, suggesting that these four planets are in a 1:2:5:7 chain of mean motion resonances. Although the orbital periods suggest this resonance, as do many follow-up studies \citep[e.g., ][]{Lillo-Box2020}, the dynamics of this system have yet to be explored. 

Following the methods of \citet{MacDonald_2022}, we seek to understand the dynamics of these systems by running \textit{N}-body simulations using the python module \texttt{REBOUND} \citep{rebound}. We run a suite of 1000 simulations, drawing initial values for planetary masses, inclinations, and orbital periods from independent, normal distributions that are centered on values constrained by current photometry. For Kepler-226, Kepler-254, and Kepler-363, we use the results from \citet{koi} for all parameters except planetary radii, for which we use the updated stellar, and therefore planetary, radii from \citet{berger2018kepler}. For Kepler-1542, we use parameters from \citet{morton2016kepler}, and for K2-32 we use the values from \citet{Heller2019}. For planets without mass constraints, we draw masses from the mass-radius relationship described in \citet{MASS_RADIUS_RELATION}\footnote{We explore a large range of masses for each planet and use the resulting resonances to constrain the planet masses. We therefore are not sensitive to any specific mass-radius relationship.}. Each simulation therefore initializes with a set of parameters that is unique from other simulations but consistent with current data.  Using the WHFast integrator \citep{rein2015}, we integrate the modeled systems for 10~Myr with a timestep of 5\% the innermost planet's period. We summarize the simulation initial conditions for our  simulations in Table~\ref{tab:sim_params}.

\begin{deluxetable*}{lcccc}
\renewcommand{\arraystretch}{0.75}
\tablecolumns{5}
\tablewidth{0pt}
\tabletypesize{\footnotesize}
\tablecaption{ Planetary Properties for Determining Resonance \label{tab:sim_params}}
\tablehead{
\colhead{Kepler-226} &
\colhead{b} &
\colhead{c} &
\colhead{d} &
\colhead{}
}
\startdata
$P$ [d] & $3.940997\pm 0.000020$ & $5.349555\pm0.000014$ & $8.109044\pm0.000094$ & \\
t$_0$ [d] & $69.09337$ & $104.80599$ & $65.80333$ & \\
$i$ [$\degree$] & 88.88$\pm$0.2 & 89.62$\pm$0.2 & 89.92$\pm$0.2 & \\
$M_p$ [$M_{\oplus}$]& $ 4.271^{+1.933}_{-1.825} {}^*$ & $ 6.237^{+2.071}_{-1.952} {}^*$ & $ 2.440^{+1.985}_{-1.244} {}^*$ &  \\
\hline
\hline
Kepler-254 & b & c & d & \\
\hline
$P$ [d] & $5.82666\pm0.00001$ & $12.41218\pm0.00008$ & $18.7464\pm0.0001$ & \\
t$_0$ [d] & $106.01$ & $75.54$ & $80.13$ & \\
$i$ [$\degree$] & 89.88$\pm$0.2 & 89.95$\pm$0.2 & 89.11$\pm$0.2 & \\
$M_p$ [$M_{\oplus}$]& 8.84$^{+2.02}_{-1.94} {}^*$ & 5.75$^{+1.99}_{-2.05} {}^*$ & 6.72$^{+2.03}_{-1.98} {}^*$ & \\
\hline
\hline
Kepler-363 & b & c & d & \\
\hline
$P$ [d] & $3.61460279\pm0.00003$ & $7.54235832\pm0.00004$ & $11.93205399\pm0.00005$ & \\
t$_0$ [d] & $67.695$ & $245965.961$ & $245975.106$ & \\
$i$ [$\degree$] & 86.02$\pm$0.2 & 88.44$\pm$0.2 & 89.52$\pm$0.2 & \\
$M_p$ [$M_{\oplus}$]& 3.05$^{+1.83}_{-1.65} {}^*$ & 4.67$^{+2.12}_{-1.90}{}^*$ & 5.34$^{+2.06}_{-1.94} {}^*$ & \\
\hline
\hline
Kepler-1542 & c & b & e & d\\
\hline
$P$ [d] & $2.8922302\pm1.472e-05$ & $3.95116882\pm1.633e-05$ & $5.10115756\pm2.409e-05$ & $5.99273738\pm2.26e-05$\\
t$_0$ [d] & $65.86465$ & $67.22178$ & $65.42378$ & $64.74864$\\
$i$ [$\degree$] & 89.89$\pm$0.2 & 88.05$\pm$0.2 & 89.68$\pm$0.2 & 88.08$\pm$0.2 \\
$M_p$ [$M_{\oplus}$]& $ 0.429^{+0.386}_{-0.228}{}^*$  & $ 0.803^{+0.823}_{-0.420}{}^*$& $ 0.850^{+0.801}_{-0.445}{}^*$ & $ 1.083^{+0.979}_{-0.570}{}^*$ \\
\hline
\hline
K2-32 & e & b & c & d\\
\hline
$P$ [d] & $ 4.34882^{+0.00069}_{-0.00075}$ & $ 8.991828^{0.000083+}_{-0.000084}$ & $ 20.66186^{+0.00102}_{-0.00098}$ & $ 31.7142^{+0.0011}_{-0.0010}$  \\
t$_0$ [d] & $1998.886$ & $ 2000.92713$ & $ 1999.42271$ & $ 2003.7913$ \\
$i$ [$\degree$] & 90.0$^{**}$ & $89.1\pm0.7$ & $89.3\pm0.9$ & $89.3\pm0.9$ \\
$M_p$ [$M_{\oplus}$]& $  1.095^{+2.248}_{-0.625}{}^*$ & $ 16.5^{+2.7}_{-2.7}$ & $ < 12.1$ & $ 10.3^{+4.8}_{-4.3}$ 
\enddata
\tablecomments{Initial conditions used for the simulations, including orbital period $P$, mid-transit time $t_0$, sky-plane inclination $i$, and planetary mass $M_p$. We initialize all planets on circular orbits. We use the values published by \citet{rowe} for all parameters of Kepler-226, Kepler-254, and Kepler-363, except for planetary radii, where we use the updated stellar and therefore planetary radii from \citet{berger2018kepler}. For Kepler-1542, we use parameters from \citet{morton2016kepler}, and for K2-32 we use the values from \citet{Heller2019}. We assume stellar masses of $0.831~M_{\odot}$ \citep{koi}, $0.943~M_{\odot}$ \citep{berger2018kepler}, $1.173~M_{\odot}$ \citep{koi}, $0.933~M_{\odot}$ \citep{koi}, and  $0.856~M_{\odot}$ \citep{Heller2019} for the stars as ordered in the table. All parameters were drawn from independent, normal distributions, centered on the nominal values with widths equal to the value's uncertainty; for parameters with unequal upper and lower uncertainties, we take the larger uncertainty as the width.\\ $^*$ planetary masses were drawn from the mass-radius relation \citet{MASS_RADIUS_RELATION}. \\ $^{**}$ At the time of this work, no estimate existed for this value, so we fix the parameter and do not draw it from a normal distribution.}
\end{deluxetable*}


\section{Results}\label{sec:Results}
For each of our five systems of interest, we run a suite of 1000 \textit{N}-body simulations for 10~Myr and analyze the results of each suite for two-body and three-body resonances. We stop integrations when any planet experiences a close encounter, defined by a distance of less than three Hill radii. To confirm a chain of resonances, we search for simulations where the three-body angle is librating or where both of the two-body angles are librating. 

We find it unlikely that Kepler-1542 and K2-32 contain any resonant chains; for each of these systems, no three-body angle librated in our simulations, regardless of planetary mass. In Kepler-1542, the resonant angle $\Theta_{e,d}=7\lambda_d-6\lambda_e-\omega_e$ librated in 82\% of simulations, and in K2-32 the resonant angles $\Theta_{c,d}=3\lambda_d-2\lambda_c-\omega_c$ and $\Theta_{e,b}=2\lambda_b-1\lambda_e-\omega_e$ librated in 70\% and 68\% of simulations, respectively. Because not all solutions to our current data lead to these angles librating, we cannot claim the planets are in resonance. 

In Kepler-226, we find that the two-body angle $\Theta'_{c,d}=3\lambda_d-2\lambda_c-\omega_d$ librates about 180$\degree$ in 99.8\% of our simulations, but with large libration amplitudes of 90.5$^{+23.19}_{15.22}$. The two-body angle $4\lambda_c-3\lambda_b-\omega_c$ librates in 42\% of our simulations, and the three-body angle circulates in all simulations. While we are therefore able to confirm the 3:2 resonance between Kepler-226c and Kepler-226d, we are not able to confirm a resonant chain.

We focus the rest of this work on the two remaining systems, Kepler-254 and Kepler-363. We summarize the results of the resonance analysis for all systems in Table~\ref{tab:res_results}.

\begin{deluxetable*}{lccc}
\renewcommand{\arraystretch}{0.75}
\tablecolumns{4}
\tablewidth{0pt}
\tabletypesize{}
\tablecaption{ Resonance Results \label{tab:res_results}}
\tablehead{
\colhead{Angle} &
\colhead{\% librating} &
\colhead{Center [$\degree$]} &
\colhead{Amplitude [$\degree$]}
}
\startdata
K2-32	&	stable = 984	&	resonant = 664		&		\\
$\Theta_{e,b}=2\lambda_b-\lambda_e-\omega_e$ &		67.58\%	&	-0.005	 $^{+	0.349	}_{-	0.315	}$ & 	48.4	 $^{+	23.8	}_{-	20.2	}$ \\
$\Theta_{b,c}=2\lambda_c-\lambda_b-\omega_b$ &		14.43\%	&	0.036	 $^{+	0.501	}_{-	0.513	}$ & 	58.3	 $^{+	14.0	}_{-	31.8	}$ \\
$\Theta_{c,d}=3\lambda_d-2\lambda_c-\omega_c$ &		69.92\%	&	0.015	 $^{+	2.073	}_{-	2.065	}$ & 	64.5	 $^{+	9.7	}_{-	18.9	}$ \\
$\Theta'_{e,b}=2\lambda_b-\lambda_e-\omega_b$ &		0.90\%	&	-5.38$^{+12.75}_{- 14.09}$ & 	134.64$^{+6.23}_{-4.81}$ \\
$\Theta'_{b,c}=2\lambda_c-\lambda_b-\omega_c$ &		0.00\%	&	\nodata & 	\nodata \\
$\Theta'_{c,d}=3\lambda_d-2\lambda_c-\omega_d$ &		6.80\%	&	-0.06$^{+15.61}_{-12.71}$ & 	\nodata \\
\hline		
Kepler-226	&	stable = 998	&	resonant = 457		&			\\
$\Theta_{b,c}=\textbf{4}\lambda_c-\textbf{3}\lambda_b-\omega_c$ &		42.00\%	&	-0.052	 $^{+	0.571	}_{-	0.504	}$ & 	119.1	 $^{+	22.2	}_{-	20.9	}$ \\
$\Theta_{c,d}=3\lambda_d-2\lambda_c-\omega_c$ &		45.80\%	&	-0.05	 $^{+	0.91	}_{-	0.88	}$ & 	135.03	 $^{+	11.8	}_{-	30.3	}$ \\
$\Theta'_{b,c}=\textbf{4}\lambda_c-\textbf{3}\lambda_b-\omega_b$ &		41.60\%	&	179.9	 $^{+	0.577	}_{-	0.428	}$ & 	137.25	 $^{+	9.21	}_{-	11.85	}$ \\
$\Theta'_{c,d}=3\lambda_d-2\lambda_c-\omega_d$ &		99.8\%	&	179.9	 $^{+	1.38	}_{-	1.11	}$ & 	90.5	 $^{+	23.19	}_{-	15.22	}$ \\
\hline	
Kepler-254	&	stable = 996	&	resonant = 422		&		\\
$\Theta_{b,c}=2\lambda_c-\lambda_b-\omega_c$ &		42.40\%	&	0.021	 $^{+	0.35	}_{-	0.39	}$ & 	118.6	 $^{+	21.48	}_{-	49.9	}$ \\
$\Theta_{c,d}=3\lambda_d-2\lambda_c-\omega_c$ &		99.20\%	&	-0.15	 $^{+	2.41	}_{-	2.22	}$ & 	$65.1^{+ 4.6} _{- 5.0}$ \\
$\Theta'_{b,c}=2\lambda_c-\lambda_b-\omega_b$ &		0.00\%	&	\nodata	 & 	\nodata \\
$\Theta'_{c,d}=3\lambda_d-2\lambda_c-\omega_d$ &		100.0\%	&	179.98	 $^{+	1.79	}_{-	1.81	}$ & 	87.1	 $^{+	12.34	}_{-	14.21	}$ \\
\hline		
Kepler-363	&	stable = 998	&	resonant = 924		&		\\
$\Theta_{b,c}=2\lambda_c-\lambda_b-\omega_c$ &		99.2\%	&	0.0029	 $^{+	0.224	}_{-	0.243	}$ & 	35.1$^{+30.0}_{-17.8}$ \\
$\Theta_{c,d}=3\lambda_d-2\lambda_c-\omega_c$ &		92.59\%	&	-0.02	 $^{+	0.54	}_{-	0.44	}$ & 	55.1	 $^{+	13.9	}_{-	13.7	}$ \\
$\Theta'_{b,c}=2\lambda_c-\lambda_b-\omega_b$ &		0.0\%	&	\nodata	  & 	\nodata  \\
$\Theta'_{c,d}=3\lambda_d-2\lambda_c-\omega_d$ &		98.8\%	&	179.97	 $^{+	0.41	}_{-	0.37	}$ & 	96.98$^{+34.82}_{-35.54}$ \\
\hline		
Kepler-1542	&	stable = 897	&	resonant = 0	&		\\
$\Theta_{c,b}=4\lambda_b-3\lambda_c-\omega_b$ &		9.81\%	&	0.16	 $^{+	0.66	}_{-	0.67	}$ & 	64.5	 $^{+	8.7	}_{-	24.3	}$ \\
$\Theta_{b,e}=5\lambda_e-4\lambda_b-\omega_b$ &		5.13\%	&	-0.17	 $^{+	1.30	}_{-	0.88	}$ & 	74.1	 $^{+	2.9	}_{-	7.4	}$ \\
$\Theta_{e,d}=7\lambda_d-6\lambda_e-\omega_e$ &		81.94\%	&	-0.05	 $^{+	0.88	}_{-	0.81	}$ & 	61.2	 $^{+	10.8	}_{-	16.5	}$ \\
$\Theta'_{c,b}=4\lambda_b-3\lambda_c-\omega_c$ &		0.50\%	&	 -5.19$^{+3.04}_{-0.85}$ & 	132.50$^{+3.34}_{-1.50}$ \\
$\Theta'_{b,e}=5\lambda_e-4\lambda_b-\omega_e$ &		2.80\%	&	-2.47$^{+12.97}_{-8.74}$ & 	127.21$^{+3.25}_{- 2.12}$ \\
$\Theta'_{e,d}=7\lambda_d-6\lambda_e-\omega_d$ &		29.2\%	&	1.08$^{+11.11}_{-13.93}$ & 	131.72$^{+7.99}_{-4.75}$ 
\enddata
\tablecomments{For each system, the number of simulations out of 1000 that survived 10~Myr, the number of simulations where all planets participate in the chain, then, for each angle, the percentage of simulations where the angle librates and the center and amplitude of the libration. For each system, all three-body angles were circulating.}
\end{deluxetable*}

\subsection{Kepler-254}\label{sec:k254-res}
\begin{figure*}[!ht]
    \centering
    \includegraphics[width=0.48\textwidth]{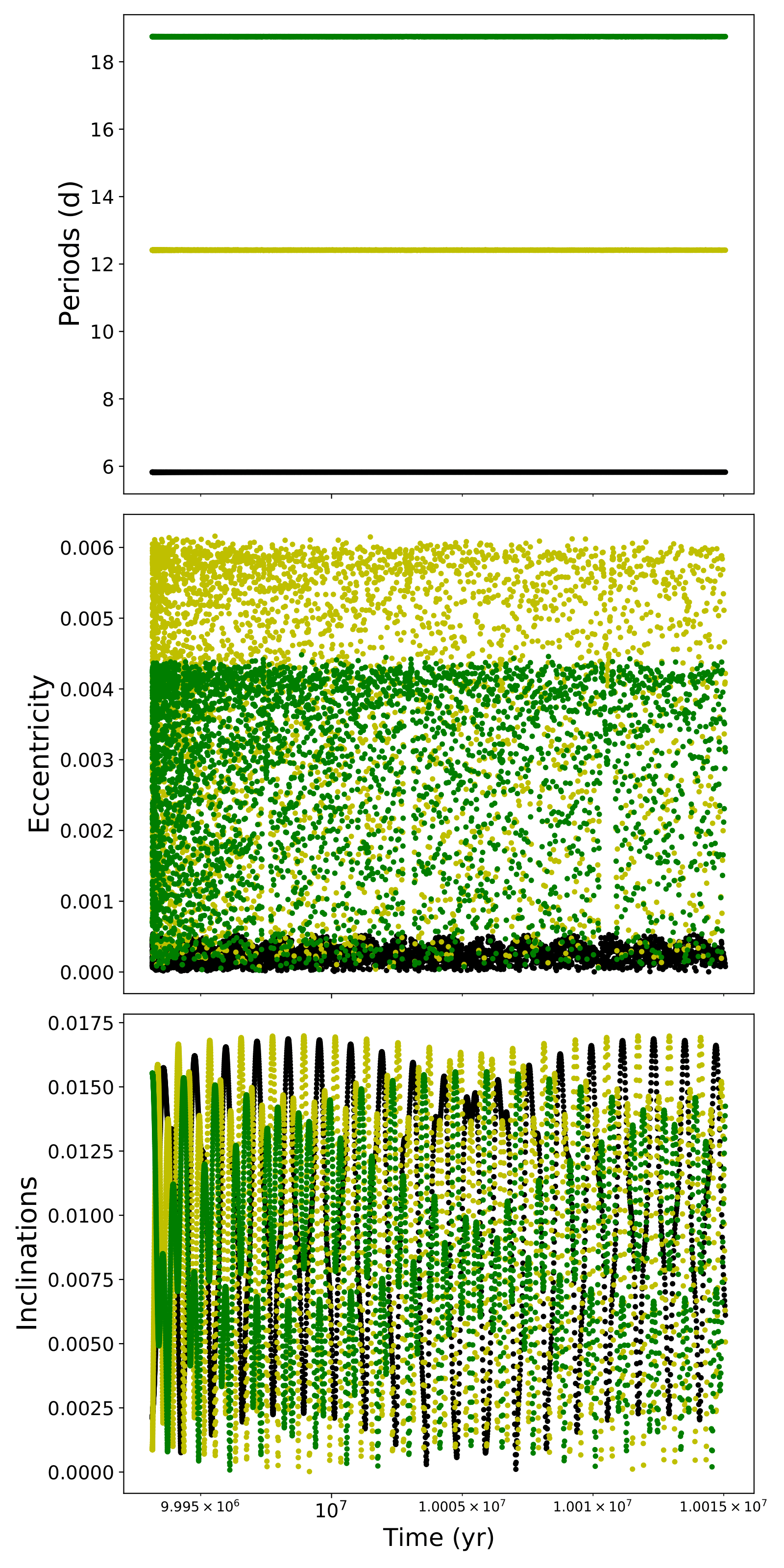}
    \includegraphics[width=0.48\textwidth]{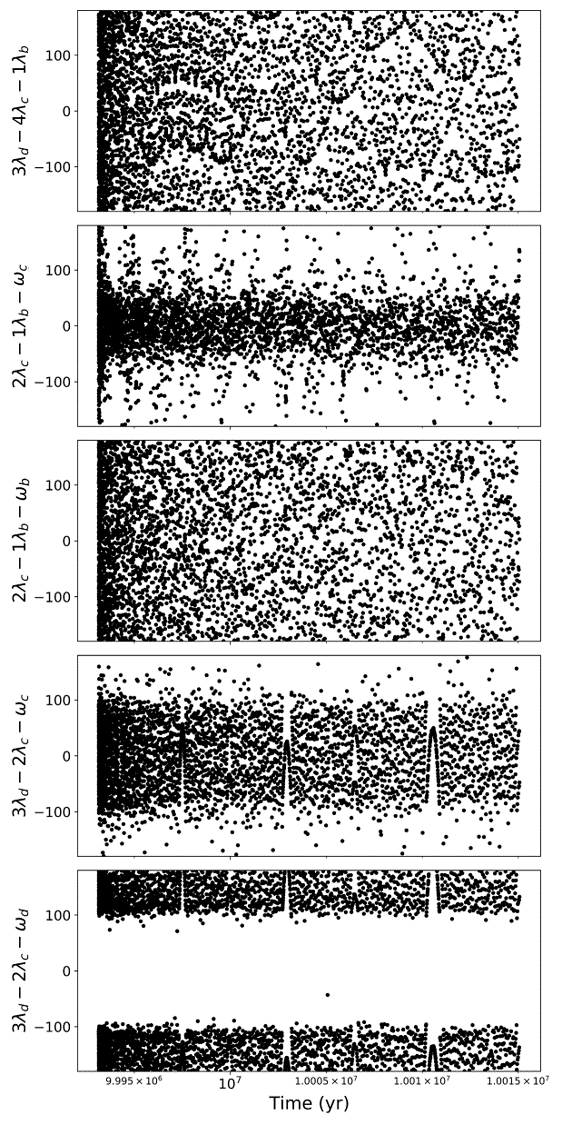}
    \caption{Example evolution of the orbital periods, eccentricities, inclinations, all four two-body resonant angles, and the three-body resonant angle of the three planets of Kepler-254. We find that the two-body angle $\Theta_{c,d}$ librates in nearly all of our simulations, the two-body angle $\Theta_{b,c}$ only librates in approximately 40\%, and the corresponding three-body angle circulates in each one. The initial values for this simulation were drawn from independent, normal distributions, as described in Section~\ref{sec:background} and summarized in Table~\ref{tab:sim_params}. We integrate this simulation beyond 10~Myr for visualization purposes.}
    \label{fig:254res}
\end{figure*}

Through our analysis, we find that nearly all (99.6\%) simulations of Kepler-254 remained stable during the 10~Myr integrations, i.e. no planets experienced a close encounter or were ejected, regardless of initial parameter values. Of these simulations, 42.4\% result in a 1:2:3 three-body resonant chain. The two-body angle $\Theta_{b,c}=2\lambda_c-\lambda_b-\omega_c$ librates in 42.4\% of the simulations, and the two-body angle $\Theta_{c,d}=3\lambda_d-2\lambda_c-\omega_c$ librates in 99.2\% of the simulations. The three-body angle $\phi_1=3\lambda_d-4\lambda_c+\lambda_b$ circulated in all of the simulations. We show the evolution of one the \textit{N}-body simulations in Figure~\ref{fig:254res}.
Given these results, we are therefore able to confirm a two-body resonance between Kepler-254c and Kepler-254d where the angle $\Theta_{c,d}$ librates around 0$\degree$ with an amplitude of $65.1^{+ 4.6} _{- 5.0}$. A three-body resonant chain is probable but requires further analysis and more precise orbits to confirm. The system could therefore benefit from follow-up observation and analysis.

\subsection{Kepler-363}\label{sec:k363-res}
Regardless of the initial parameters, nearly all 1000 simulations of Kepler-363 remained stable for the 10~Myr integration. We find the 2:1 resonant angle $\Theta_{b,c}=2\lambda_c-\lambda_b-\omega_c$ librates in 99.2\% of simulations, and the 3:2 resonant angle $\Theta_{c,d}=3\lambda_d-2\lambda_c-\omega_c$ librates in 92.6\% of simulations. Of all 1000 simulations, 92.4\% result in a three-body 1:2:3 resonant chain. 

\begin{figure*}[!ht]
    \centering
    \includegraphics[width=0.48\textwidth]{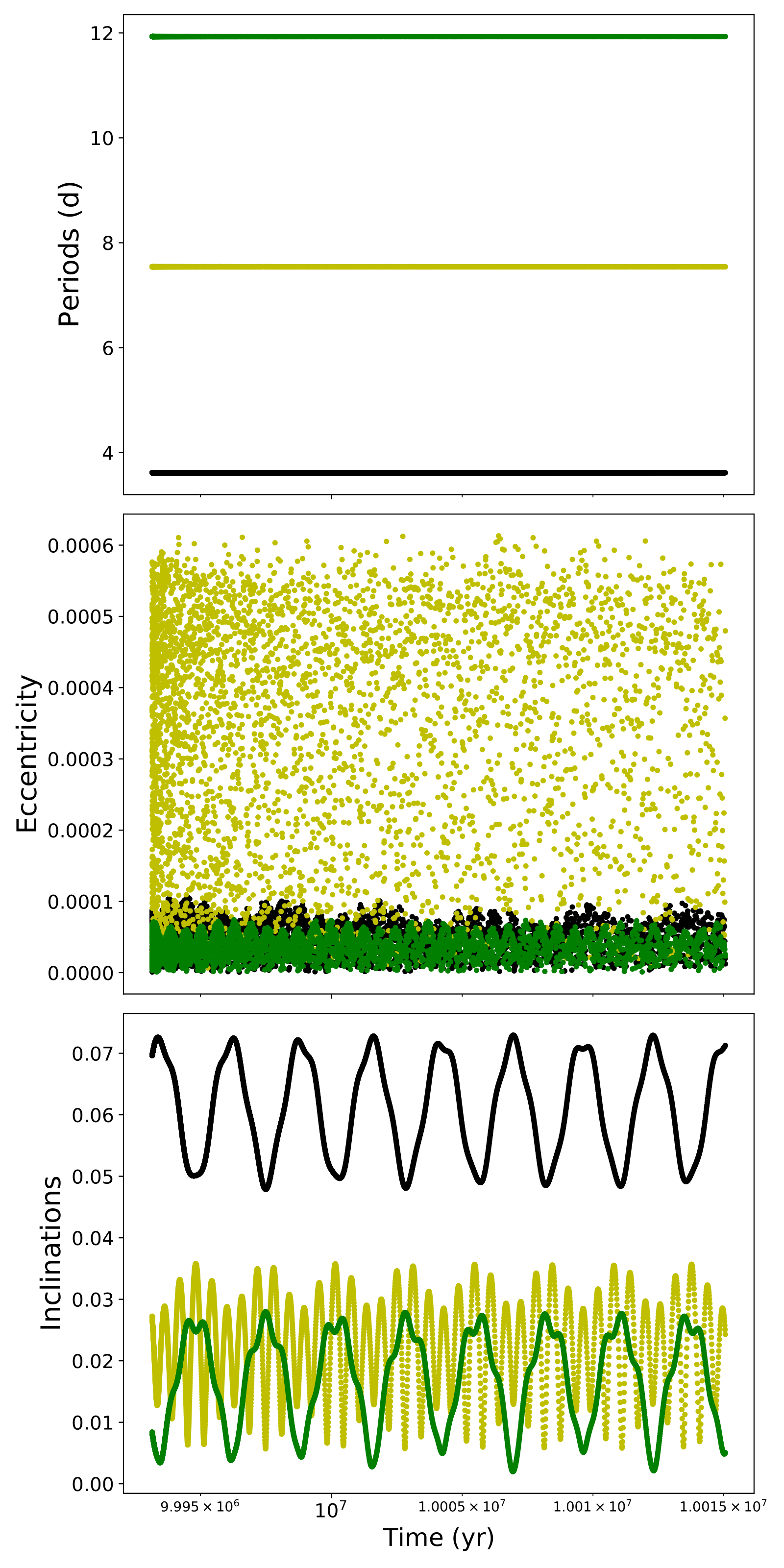}
    \includegraphics[width=0.48\textwidth]{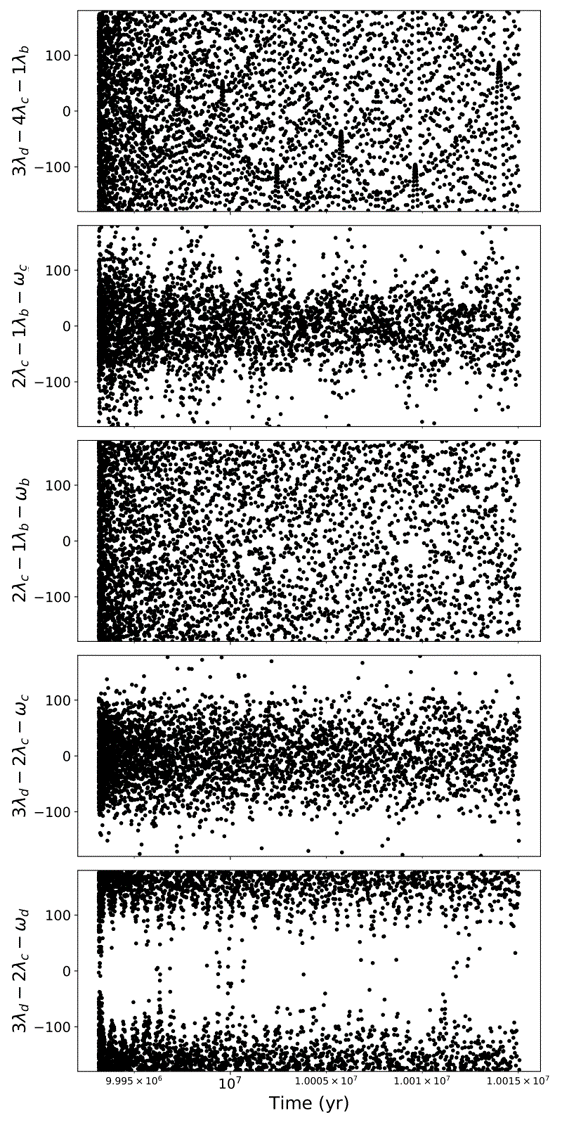}
    \caption{Example evolution of the orbital periods, eccentricities, inclinations, all four two-body resonant angles, and the three-body resonant angle of the three planets of Kepler-363. We find that the two-body angles $\Theta_{b,c}$, $\Theta_{c,d}$, and $\Theta'_{c,d}$, librate in nearly all of our simulations, but the corresponding three-body angle circulates in each one. The initial values for this simulation were drawn from independent, normal distributions, as described in Section~\ref{sec:background} and summarized in Table~\ref{tab:sim_params}. We integrate this simulation beyond 10~Myr for visualization purposes.}
    \label{fig:363res}
\end{figure*}

The two-body angles $\Theta_{b,c}$ and $\Theta_{c,d}$ librate about 0$\degree$ with moderate amplitudes of 35.1$^{+30.0}_{-17.8}$ and 55.1$^{+13.9}_{-13.7}$, respectively, and the two-body angle and $\Theta'_{c,d}$  librates around 180$\degree$ with large amplitudes of 96.98$^{+34.82}_{-35.54}$. Curious enough, the three-body angle $\phi = 3\lambda_d-4\lambda_c+\lambda_b$ does not librate in any of our simulations. We discuss the implications of this circulating angle in more detail in Section~\ref{sec:discussion}. We show the evolution of one the \textit{N}-body simulations in Figure~\ref{fig:363res}.

\section{Discussion}\label{sec:discussion}
With the confirmation of resonance, we are able to study additional information about a system and its planets. In particular, resonances allow us to constrain planetary masses and orbits and to explore the formation and subsequent dynamical history of the planets.

\subsection{Using Resonance to Constrain Masses and Orbits}\label{sec:constraints}

We explore the differences in planetary parameters between simulations that resulted in resonance and those that did not. We perform a two-sample Kolmogorov-Smirnov test, exploring the null hypotheses that the masses, eccentricities, and orbital periods of the planets in resonance and the planets not in resonance are drawn from the same distribution. As an example, we take the distribution of masses of Kepler-363b from simulations where $\Theta_{b,c}$ librates as one sample for the K-S test, and the distribution of that planet's mass from simulations where the same angle circulates as the second sample.

For all parameters except the eccentricity of Kepler-363c, we recover large \textit{p}-values (p$~>~$0.05) and fail to reject the null hypothesis. For Kepler-363c's eccentricity, we recover a \textit{p}-value of 0.018, suggesting that the two distributions are statistically different. We find that the resulting eccentricity for simulations with a librating $\Theta_{c,d}$ is smaller than for those with a circulating $\Theta_{c,d}$ (2.3$^{+1.8}_{-1.4}\times10^{-4}$ and 3.0$^{+1.5}_{-1.7}\times10^{-4}$, respectively).  Although we are not able to use the system's resonances to constrain the planets' masses, we do find that this system's resonant state is not very dependant on the planetary masses, confirming that more precise mass measurements are not necessary to confirm these resonances.

\subsection{Constraining dynamical history}\label{sec:form}

With confirmed resonances, we are now able to study each system's formation and evolution. Although resonant chains are typically seen as the hallmark of disk-driven migration, two additional pathways exist to form resonant chains that are each consistent with in situ formation \citep{macdonald2018three}. Following the prescription of \citet{macdonald2018three}, the three chain formation pathways are long-scale migration (\textbf{LM}; hypothesizes the planets were formed both further from their star and each other when compared to current observations), short-scale migration (\textbf{SM}; planets formed near current observations, just outside of resonance, where small shifts in the planets' semi-major axes will lead to resonance), and eccentricity dampening (\textbf{ECC}; planets formed near current observations, just outside of resonance, where damping to the planets' eccentricities will lead to resonance). 

To study the formation of the resonances in Kepler-254 and Kepler-363, we follow the methods of \citet{macdonald2018three} which we briefly describe here. For each formation pathway, we run a suite of 500 \textit{N}-body simulations with the same initial conditions shown in Table~\ref{tab:sim_params} except with inflated orbital periods. We use the \texttt{modify\_orbits\_forces} routine in the REBOUNDx library \citep{reboundx} and the WHFast integrator \citep{rein2015}. For the \textbf{LM} simulations, we initialize the inner planet at 1 au from its host star and start the other planets just wide of the observed resonances\footnote{We round each planet's semimajor axis up to the nearest 0.1 au to ensure the planets start out of resonance.}. For the \textbf{SM} and \textbf{ECC} simulations, we initialize the planets a small percentage wide of their observed orbits, where we draw this percentage for each planet and each simulation from a normal distribution of $N[5,3]$\%. All simulations start with the planets out of resonance. We then form the resonant chains by damping the semi-major axes and/or eccentricities of the planets, following the prescription in \citet{papaloizou2000}. For the \textbf{LM} and \textbf{SM} simulations, we damp only the outer planet's eccentricity and semi-major axis\footnote{Both the direction and rate of migration for each planet will depend on conditions of the disk and are therefore unknown. By simulating the migration of only the outer planet, we implicitly assume that the migration timescale of the inner planets is much longer.}, and for the \textbf{ECC} simulations, we damp the eccentricity of all planets. We draw the timescales for the semi-major axis damping ($\tau_a$) and eccentricity damping ($\tau_e$) from independent, log-uniform distributions of log~$\tau_a$ = U[7, 9] yr, log~$\tau_e$ = U[4, 6] yr; log~$\tau_a$ = U[6, 9] yr, log~$\tau_e$ = U[4, 7] yr; and log~$\tau_e$ = U[5, 7] yr for the \textbf{LM}, \textbf{SM}, and \textbf{ECC} suites, respectively. We explore a wide range of damping timescales, representing a wide range of disk conditions, to avoid fine-tuning our simulations.

We integrate each system forward with a timestep of 5\% the innermost planet's observed orbital period. After 5$\times10^6$ years, we ``turn off'' the damping effects and integrate for another 0.25 Myr to ensure stability after the gas disk would dissipate. We then study each resulting simulation for librating two- and three-body resonant angles. 

\begin{figure*}[!ht]
    \centering
    \includegraphics[width=0.8\textwidth]{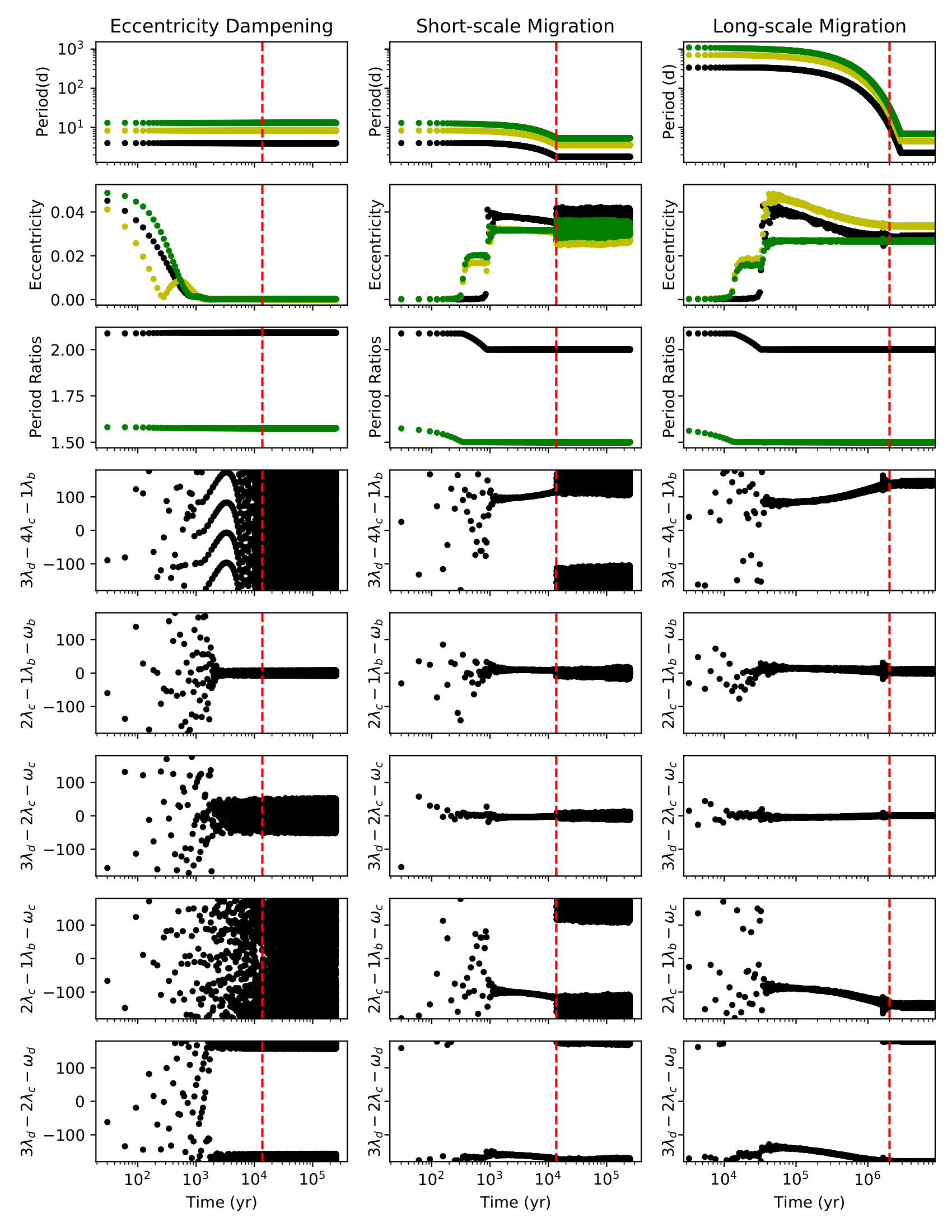}
    \caption{Example evolution of systems like Kepler-363, forming the resonant chain through three formation pathways: eccentricity damping only, short-scale migration, and long-scale migration. The period ratio marked as black dots is the ratio between planets b and c, the period ratio marked as green dots is the ratio between planets c and d, and the vertical red line indicates when we ``turn-off'' the damping effects. Although each pathway is able to lock the planets into both two-body resonances, both short-scale migration and long-scale migration result in the libration of the three-body angle $3\lambda_d-4\lambda_c-\lambda_b$ which we find to be circulating.}
    \label{fig:evo}
\end{figure*}

\begin{figure*}[!ht]
    \centering
    \includegraphics[width=0.8\textwidth]{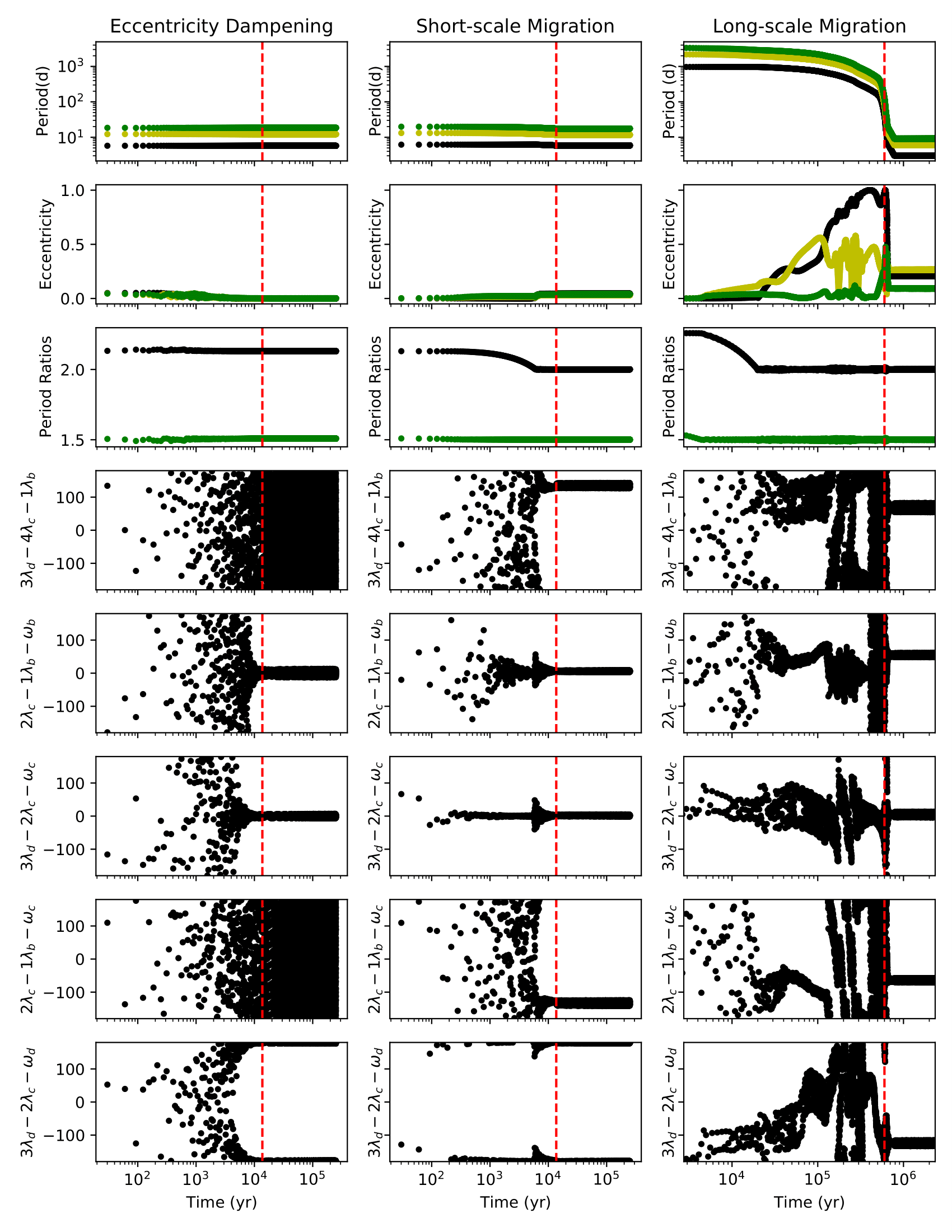}
    \caption{Example evolution of systems like Kepler-254, forming the resonant chain through three formation pathways: eccentricity damping only, short-scale migration, and long-scale migration. The period ratio marked as black dots is the ratio between planets b and c, the period ratio marked as green dots is the ratio between planets c and d, and the vertical red line indicates when we ``turn-off'' the damping effects. Both eccentricity dampening and short-scale migration are lock the planets into both two-body resonances while only short-scale migration results in the libration of the three-body angle $3\lambda_d-4\lambda_c-\lambda_b$ which we find to be circulating. Long-scale migration did not lead to enough simulations remaining stable to yield statistically significant results.}
    \label{fig:evo254}
\end{figure*}

We find we are able to produce a full three-body resonant chain in systems like Kepler-254 and Kepler-363 through all three formation pathways. However, each formation pathway yields unique results, which we discuss in turn below. We summarize the centers and amplitudes of librating angles resulting from each formation pathway in Table~\ref{tab:otherres}, and we compare examples from each of these formation pathways in Figures~\ref{fig:evo} and~\ref{fig:evo254}.

\textit{Short-scale migration:} For both systems, short-scale migration (\textbf{SM} suite) results in the three-body angle $\phi=\Theta_{b,c}-\Theta_{c,d}$ librating in some of the simulations (34\% and 25\% for Kepler-254 and Kepler-363, respectively), and librating about 180$\degree$, $\sim$285$\degree$, and a third center with moderate amplitudes ($\sim10-20\degree$). 

\textit{Long-scale migration:} Since very few of the \textbf{LM} simulations for Kepler-254 remained stable for the full integration time, and with only one simulation in resonance, we are unable to perform any meaningful statistical analysis on this suite. The long-scale migration for Kepler-363 resulted in very few simulations where $\phi$ librates and only 27\% of the stable simulations with a three-body resonant chain. 

\textit{Eccentricity-damping:} We find that eccentricity-damping results in the libration of the two-body angles $\Theta_{b,c}$, $\Theta_{c,d}$, and $\Theta'_{c,d}$ for both Kepler-254 and Kepler-363 in about half of the simulations, but very rarely results in the libration of $\Theta'_{b,c}=2\lambda_c-\lambda_b-\omega_b$ or of the three-body angle $\phi$. For Kepler-254, $\Theta_{b,c}$ and $\Theta_{c,d}$ each librate about 0$\degree$ with small amplitudes of $5.96^{+5.23}_{-0.62}$ and $5.76^{+9.26}_{-1.21}$, respectively, similar to the centers we recover in Section~\ref{sec:k254-res} but with significantly smaller amplitudes.  For Kepler-363, $\Theta_{b,c}$ and $\Theta_{c,d}$ each librate about 0$\degree$ with amplitudes of $4.22^{+2.45}_{-0.47}$ and $28.32^{+4.05}_{-2.13}$, respectively, similar to the centers we recover in Section~\ref{sec:k363-res} but, again, with significantly smaller amplitudes. 

In Section~\ref{sec:k254-res}, we confirmed the two-body resonance between Kepler-254c and Kepler-254d, but we were unable to confirm a resonance between the inner planet pair in the system. Since each of the formation pathways resulted in the libration of this angle and therefore each pathway is possible given our current data, we cannot select one pathway over another as more probable. We find it likely that the resonant chain of Kepler-363 formed through eccentricity- damping, which we discuss in more detail below in Section~\ref{sec:weird3br}. 

\begin{deluxetable*}{lccc|lccc}
\renewcommand{\arraystretch}{1.0}
\tablecolumns{8}
\tablewidth{0pt}
\tabletypesize{}
\tablecaption{ Formation Pathway Results \label{tab:otherres}}
\tablehead{
\colhead{Angle} & 
\colhead{\% librating} & 
\colhead{Center [$\degree$]} & 
\colhead{Amplitude [$\degree$]} & 
\colhead{Angle} & 
\colhead{\% librating} & 
\colhead{Center [$\degree$]} & 
\colhead{Amplitude [$\degree$]}
}
\startdata
Kepler-254	& \textbf{SM} & stable = 481/500 & res = 368/500 & Kepler-363	&	\textbf{SM} & stable = 474/500 & res = 356/500 \\
$\phi_1$ & 8.11 & $180.32 ^{+ 2.43 }_{- 2.20 }$ &  $15.33 ^{+ 18.7 }_{- 7.98 }$ & $\phi_1$ & 11.6 & $90.99 ^{+ 19.23 }_{- 17.78 }$ & $13.64 ^{+ 17.38 }_{- 9.18 }$ \\
 & 17.0 &  $81.75 ^{+ 28.74 }_{- 11.25 }$ &  $14.07 ^{+ 19.00 }_{- 8.12 }$ &  & 6.3 & $ 180.37 ^{+ 10.54 }_{- 2.55 }$ & $20.86 ^{+ 30.08 }_{- 10.24 }$ \\
 & 9.1 & $287.56 ^{+ 7.94 }_{- 6.50 }$ & $22.48 ^{+ 19.64}_{- 12.91 }$ &  & 7.4 & $ 285.27 ^{+ 5.85 }_{- 15.72 }$ & $20.93 ^{+ 12.12 }_{- 12.03 }$ \\
$\Theta_{b,c}$ & 49.9 & $ 0.11 ^{+ 3.38 }_{- 0.85 }$ & $  7.95 ^{+ 37.17 }_{- 4.04 }$ & $\Theta_{b,c}$ & 8.6 &  $ -48.88^{+ 14.25 }_{- 8.56 }$ & $ 14.05 ^{+ 9.50 }_{- 7.08 } $ \\
 & 10.8 &  $  -48.62 ^{+ 11.05 }_{- 8.02 }$ & $  10.64 ^{+ 8.53 }_{- 5.50 }$ &   & 68.6 & $0.17 ^{+ 16.90 }_{- 1.06 }$ & $12.09 ^{+ 28.53 }_{- 7.99 }$ \\ 
 & 16.2 &  $  45.53 ^{+ 9.70 }_{- 10.71 }$ & $  11.48 ^{+ 8.50 }_{- 6.27 }$ &   $\Theta_{c,d}$ & 82.1 & $ 0.02 ^{+ 2.63 }_{- 1.71 } $ & $15.72^{+ 14.21}_{ - 12.57}$ \\
$\Theta_{c,d}$ & 84.0 & $0.04 ^{+ 3.40 }_{- 1.09 }  $ & $7.44^{+20.09}_{-5.34}$ &   $\Theta'_{b,c}$ & 13.1 & $279.32 ^{+ 16.34 }_{- 26.93 }$ & $8.53 ^{+ 9.41 }_{- 4.82 }$ \\
$\Theta'_{b,c}$ & 21.2 & $287.45 ^{+ 6.09 }_{- 33.69 }$ & $  7.24 ^{+ 8.55 }_{- 3.98 }$ &    & 8.4 &  $65.87^{+10.33}_{-6.30}$ & $  11.62 ^{+ 6.35}_{- 5.09 }$ \\
 & 11.6 & $66.59 ^{+ 8.83 }_{- 4.17 }$ & $  10.12 ^{+ 8.16 }_{- 5.96 }$ &  & 7.2 & $  179.04 ^{+ 3.88 }_{- 21.02 }$ & $  31.02 ^{+ 43.83 }_{- 21.05 }$ \\
 & 6.7 & $179.70 ^{+ 2.20 }_{- 2.93 }$ & $ 15.55 ^{+ 36.56 }_{- 8.64 }$ & $\Theta'_{c,d}$ & 82.9 & $179.97 ^{+ 1.42 }_{- 1.58 } $ & $ 11.30^{+ 18.78 }_{- 8.21}$ \\
$\Theta'_{c,d}$ & 81.3 & $179.99 ^{+ 14.76 }_{- 1.02} $ & $7.78^{+18.20}_{-6.24}$ &  &  &  &  \\
\hline
Kepler-254	& \textbf{ECC} & stable = 500/500 & res = 238/500 &  Kepler-363	& \textbf{ECC} & stable = 500/500 & res = 229/500 \\
$\phi_1$ & 0.0 & \nodata & \nodata  & $\phi_1$ & 0.6 &  \nodata  & \nodata  \\
$\Theta_{b,c}$ & 47.6 & $  -0.017 ^{+ 0.587 }_{- 0.541 } $ & $ 5.96 ^{ 5.23}_{- 0.62} $ & $\Theta_{b,c}$ & 46.8 & $0.019^{+0.392 }_{-0.412 }$ & $4.22^{+2.45}_{-0.47}$ \\
$\Theta_{c,d}$ & 62.0 & $  -0.013 ^{+ 0.478 }_{- 0.522 } $ & $5.76^{+ 9.26}_{ - 1.21} $ & $\Theta_{c,d}$ & 47.0 & $-0.031^{+2.242}_{-1.735}$ & $28.32^{+4.05}_{-2.13}$ \\
$\Theta'_{b,c}$ & 0.0 & \nodata & \nodata & $\Theta'_{b,c}$ & 1.4 & \nodata & \nodata \\
$\Theta'_{c,d}$ & 62.2 & $  180.00 ^{+ 0.31 }_{- 0.22 }$ & $ 3.69^{+ 10.61}_{ - 0.83} $ & $\Theta'_{c,d}$ & 47.2 & $179.99^{+0.84}_{-0.79}$ & $15.57^{+ 1.46}_{-0.89}$ \\
\hline
Kepler-254	& \textbf{LM} & stable = 13/500 & res = 1/500 & Kepler-363	& \textbf{LM} & stable = 73/500 & res = 21/500 \\
$\phi_1$ & 0.2 & \nodata & \nodata  & $\phi_1$ & 1.2 & \nodata & \nodata  \\
$\Theta_{b,c}$ & 0.2 & \nodata & \nodata & $\Theta_{b,c}$ & 6.4 & $0.28^{+ 6.81}_{- 2.38}$ & $9.16^{+ 50.69}_{- 5.05}$ \\
$\Theta_{c,d}$ & 0.2 & \nodata& \nodata & $\Theta_{c,d}$ & 6.8 & $0.19^{+ 5.94}_{- 1.98}$ & $40.42^{+ 38.66}_{ - 38.84}$ \\
$\Theta'_{b,c}$ & 0.2 & \nodata & \nodata & $\Theta'_{b,c}$ & 1.2 & \nodata & \nodata \\
$\Theta'_{c,d}$ & 0.2 & \nodata & \nodata & $\Theta'_{c,d}$ & 8.8 & $179.99^{+ 2.97}_{ - 3.84}$ & $26.93^{+ 20.19}_{ - 23.70}$ \\
\enddata
\tablecomments{For each system, the number of simulations that survived the full integration, the number of simulations where all planets participate in a 1:2:3 chain, then, for each angle, the percentage of simulations where the angle librates and the center and amplitude of the libration. We do not include center or amplitude data for angles librating in fewer than 5\% of simulations.}
\end{deluxetable*}

\subsection{Unique Dynamical Configuration}\label{sec:weird3br}

The three planets of Kepler-363 are locked in a three-body resonance, where both two-body angles librate and the three-body angle $\phi = \Theta_{c,d}-\Theta_{b,c} = 3\lambda_d-4\lambda_c+\lambda_b$ circulates; the three-body angle even circulates in most of our chain-formation simulations (see Table~\ref{tab:res_results}). Typically, the three-body angle will librate if the associated two-body angles librate\footnote{Although the opposite is not true in the case of pure three-body resonance}, and so we must ask: how could this resonant chain form \textit{without} the libration of this three-body angle? We also find that the angle $\Theta'_{b,c}=2\lambda_c-\lambda_b-\omega_b$ always circulates in our simulations and the angle $\Theta'_{c,d}=2\lambda_d-\lambda_c-\omega_d$ always librates in our simulations. We can use all five resonant angles ($\Theta_{b,c}$, $\Theta_{c,d}$, $\Theta'_{b,c}$, $\Theta'_{c,d}$, and $\phi$) to study the possible formation history of Kepler-363; a likely formation pathway would result in systems with dynamics similar to those we observe: $\Theta_{b,c}$, $\Theta_{c,d}$, and $\Theta'_{c,d}$ are librating but $\Theta'_{b,c}$ and $\phi$ are circulating. 

\textit{Short-scale migration:} The angle $\Theta'_{b,c}$ librates in 38\% of our \textbf{SM} simulations, and $\Theta'_{c,d}$ librates in 82.9\% of our \textbf{SM} simulations. In addition, the three-body angle $\phi$ librates in 33\% of our simulations. If $\Theta'_{b,c}$ and $\phi$ are indeed circulating, we find it unlikely that the resonant chain formed through short-scale migration.

\textit{Long-scale migration:} As discussed above, it is challenging to form this chain through long-scale migration as the system becomes unstable without large eccentricity damping. However, we still find numerous sets of initial parameters that result in $\phi$ librating. It is therefore possible that  that this resonant chain formed through long-scale migration but requires more fine-tuning of parameters.

\textit{Eccentricity damping:} From our 500 simulations, only seven (1.4\%) result in the libration of $\Theta'_{b,c}$, and only three (0.6\%) result in the libration of $\phi$. Of the seven simulations resulting in the libration of $\Theta'_{b,c}$, one simulation has only this angle librating and all other angles circulating, one simulation does not result in $\Theta_{c,d}$ librating, one simulation results in all angles librating, and the remaining four simulations result in all two-body angles librating. The angle $\phi$ librates in one simulation where all angles librate and in two simulations where all other angles circulate. We therefore find that it is challenging for $\Theta'_{b,c}$ and $\phi$ to librate if this chain was formed without any change in the planets' semi-major axes. 

Since we are only able to simulate the formation of resonant chains in systems \textit{similar} to Kepler-363, we caution against claims of one formation mechanism; however, we find that the angles $\Theta'_{b,c}$ and $\phi$ do not librate in chains formed with eccentricity-damping when the angles $\Theta_{b,c}$, $\Theta_{c,d}$, and $\Theta'_{c,d}$ \textit{do} librate, resulting in the dynamics we observe. Resonant chains formed through short-scale and long-scale migration both result in the libration of $\Theta'_{b,c}$ and $\phi$ in the majority of simulations where the other angles librate.


\section{Conclusion}\label{sec:conclusion}
Planets in mean motion resonance with one another periodically exchange energy and angular momentum, enabling us to constrain the formation history of individual systems and identify indicators of formation history in other systems. Because the confirmation of resonance requires an in-depth study of a system's dynamics, most resonances have not been confirmed. Here, we perform such a dynamical study of five multi-planet systems whose period ratios suggest they could be in resonance.

For each system, we run a suite of \textit{N}-body simulations, exploring the full range of possible planetary and orbital parameters as constrained by available data. We confirm that two planets are in resonance if their critical resonant angle librates in at least 90\% of our simulations. Kepler-1542 and K2-32 each contain at least one planet pair that is likely in resonance, but the uncertainties on the planet masses and orbits prohibit us from confirming these resonances. We confirm the 3:2 resonance between Kepler-226c and Kepler-226d, confirm the 3:2 resonance between Kepler-254c and Kepler-254d, and confirm the 1:2:3 resonant chain between the three planets of Kepler-363. For each of these systems, we find that the three-body critical angle $\phi = \Theta_{c,d}-\Theta_{b,c} = 3\lambda_d-4\lambda_c+\lambda_b$ circulates in all of our simulations, even when both $\Theta_{c,d}$ and $\Theta_{b,c}$ librate. All five of these systems could benefit from additional data and certainly additional analysis, as their proximity to resonance likely results in measurable TTVs.

We explore the dynamical history of Kepler-254 and Kepler-363, integrating the systems through three potential resonant chain formation pathways: long-scale migration, short-scale migration, and only eccentricity damping. Under our simple migration model, both migration pathways lead to the libration of the three-body angle, suggesting that the resonances in these two systems are more likely to have formed in the absence of migration.

Our methods to confirm or constrain resonances within systems in the absence of high-precision data can be applied to other systems with near-resonant planets and would provide a list of potential new resonances that require further analysis. With the confirmation of new resonances and particularly new resonant chains, we are able to fully leverage the benefits of resonances and constrain the formation history of exoplanetary systems.

We thank the anonymous referee for the constructive review that improved this work. The authors acknowledge use of the ELSA high performance computing cluster at The College of New Jersey for conducting the research reported in this paper. This cluster is funded in part by the National Science Foundation under grant numbers OAC-1826915 and OAC-1828163.

\bibliographystyle{aasjournal}
\bibliography{main}

\end{document}